# Graph Identification of Proteins in Tomograms (GRIP-Tomo) 2.0: Topologically Aware Classification for Proteins

Short title: GRIP-Tomo 2.0

Chengxuan Li[1], August George[2,3], Reece Neff[4], Doo Nam Kim[3], Trevor Moser[2,3], Kate Baldwin[2,3], Malio Nelson[3], Arsam Firoozfar[2,3], James E Evans[2,3, *], and Margaret S Cheung[1,3, *]

[1]*Department of Physics, University of Washington, Seattle, Washington 98195, USA*

[2]*Environmental Molecular Sciences Division, Pacific Northwest National Laboratory, Richland, WA 99354, USA*

[3]*Biological Sciences Division, Pacific Northwest National Laboratory, Richland, WA 99354, USA*

[4]*North Carolina State University, Raleigh, NC 27695, USA*

*Corresponding author: Margaret S Cheung, margaret.cheung@pnnl.gov and James E Evans, James.Evans@pnnl.gov*

# Abstract

Cryo-electron tomography (cryo-ET) enables structural characterization of biomolecules under near-native conditions. Existing approaches for interpreting the resulting three-dimensional volumes are computationally expensive and have difficulty interpreting density associated with small proteins/complexes. To explore alternate approaches for identifying proteins in cryo-ET data we pursued a Graph Network and topologically invariant approach. Here, we report on a fast algorithm that distinguishes volumes containing protein density from noise by searching for nuances of evolutionarily conversed motifs and the geometrical characteristics of protein structure. GRIP-Tomo 2.0 is a machine-learning pipeline that extracts interpretable topological features of protein structures within noisy experimental backgrounds. Compared to version 1.0, the new pipeline includes three upgrades that significantly improve performance including synthetic tomogram generation simulating realistic noise, graph-based persistent feature extraction as protein fingerprints, and high-performance computing acceleration. GRIP-Tomo 2.0 achieves over 90% accuracy in distinguishing proteins from noise for synthetic datasets and over 80% accuracy for real datasets, which represents a foundational step toward advancing cryo-ET workflows and empowering automated detection of both small and large proteins for visual proteomics.

Teaser: A fast algorithm distinguishes protein density from noise by searching for nuances of structural motifs

# Introduction

A living cell contains dynamic, spatially complex protein assemblies that drive cellular processes and are highly sensitive to its functional state [1, 2]. Characterization of protein assemblies in their native environment is essential for relating their native structural properties to the phenotype of a cellular state [3-5]. However, characterizing the three-dimensional (3D) structures of these protein assemblies *in situ* is challenging due to the highly complex intracellular environment. Since the "resolution-revolution"[6] in the field of cryogenic electron microscopy (cryo-EM), cryogenic electron tomography (cryo-ET) of whole cells or sections has proven capable of resolving the structures of complex cellular architectures *in situ* [3, 7]. However, achieving high quality structures from *in situ* environments using cryo-ET remains a challenging prospect due in part to the inherently poor signal to noise ratio necessarily found in cryo-ET data. The most common pipelines for determining protein structures from noisy cryo-ET data involves picking[8, 9] and averaging many copies of a structurally homogenous protein in order to amplify signal – an approach commonly referred to as subtomogram averaging (STA)[10-12]. While STA pipelines have been demonstrated to solve protein structures from *in situ* cellular environments, there remain difficulties in recovering structures that are small in size or structurally heterogeneous. As a result, there is a need to explore other computational methods for signal extraction from noisy cryo-ET data in unlocking the full potential of the entire tomogram[13].

To overcome these challenges, we sought to develop a principled algorithm that identifies protein structures according to the nuances of evolutionarily conversed motifs [14], the arrangement of segments in a protein structure, as high-value information against noises. Inspired by the notion that only a small fraction of unique, evolutionarily conserved protein structures are determined from billions of known sequences [15] [16], we built a fast algorithm, Graph Identification of Proteins in Tomograms (GRIP-Tomo) [17], to identify protein densities in a subtomogram from background noise by seeking the unique features of evolutionarily conserved motifs in a single pass. GRIP-Tomo transforms protein structures into mathematical graphs and distinguishes them from each other according to the topologically invariant features represented by a graph. Our previous version of GRIP-Tomo, referred to as "GRIP-Tomo 1.0", demonstrated proof-of-principle success for protein identification among synthesized single particle sub-tomograms without realistic background noise. We note that while other works have incorporated topological analysis for protein structures[18-20] or membrane-bound complexes in cryo-ET [21], to our knowledge GRIP-Tomo is the first method to apply topological data analysis for distinguishing protein densities according to their unique shapes and patterns in sub-tomograms in a single pass.

In this follow up work, we validated GRIP-Tomo 2.0 by including realistic background noise in the synthesized sub-tomograms in order to train the algorithm to differentiate whether protein signals are present in a volume in a single pass. This program incorporates topologically aware identification of protein density from noise by incorporating explainable machine learning and cross-domain classification [22-25] that transfers the knowledge of high-value features from evolutionarily conserved motifs in a protein structure. It includes the module of synthesizing 'mock' sub-tomograms with realistic noise and tunable imaging artifacts such as dosage per tilt ($D$) and sample thickness ($z$).

GRIP-Tomo 2.0 deploys topological data analysis (TDA) [26-30], leveraging persistent topological features in spatial scales as structural 'fingerprints' to comprehensively describe the shapes as well as the embedded holes or cavities within hierarchical structures of a macromolecular complex. We deployed the pipeline on high-performance computing (HPC) platforms to efficiently compute the persistent features. We included a new module for interpretable machine-learning to classify protein densities from noises in cryo-ET data. Through a systematic examination of imaging parameters, including sample thickness ($z$) and dosage per tilt ($D$), we demonstrate that graph-theoretic features depend on $z$ and $D$, inform data calibration for cross-domain learning from the density maps of correlative protein mixture samples in this work, and enable interpretation for data curation. These curated datasets from synthesized and comparable experimental density maps provide the sandbox to scrutinize topological features of protein structures as structural fingerprints to guide users for accurately distinguishing protein density from noise, laying foundation for expanding GRIP-Tomo into more complex dataset environments in a single pass.

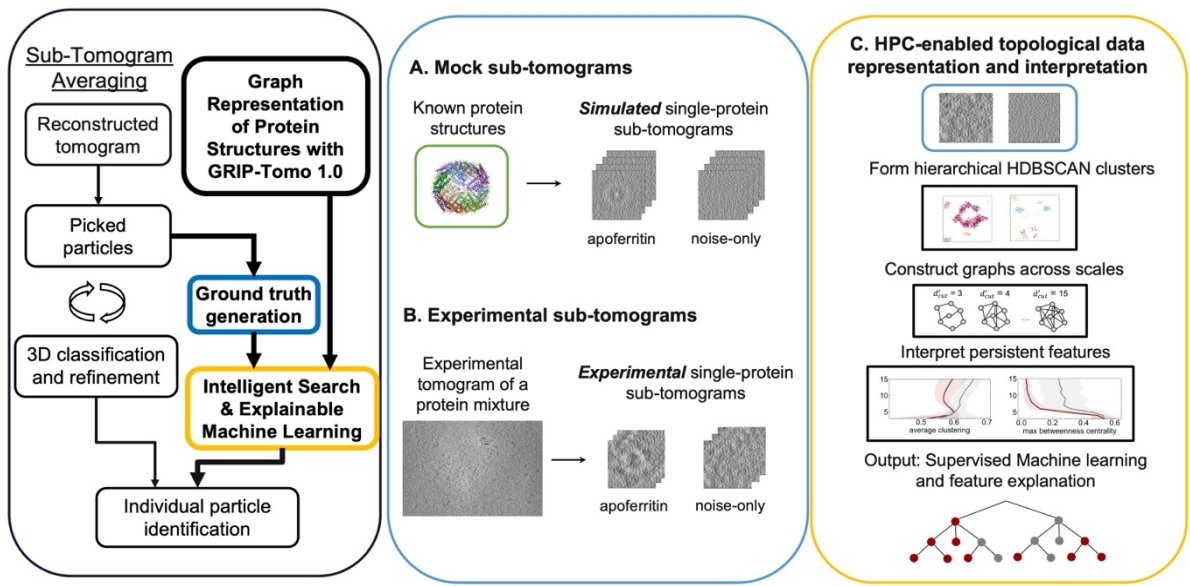

**Figure 1. GRIP-Tomo 2.0 framework for interpretable identification of proteins in cryo-ET volumes in a single pass.** Overview of the GRIP-Tomo 2.0 pipeline, designed to improve the identification of protein density from noise in sub-tomogram averaging (STA) through explainable machine learning with topologically aware feature extraction. The pipeline involves **(A)** generating reliable ground truth using mock and **(B)** correlative

experimental data for cross-domain classification, and **(C)** enabling interpretable machine learning workflows.

# Results

1. **An Overview of the GRIP-Tomo 2.0 Pipeline for Distinguishing Protein Density from Noise in Cryo-ET Volumes in a Single Pass**

*Overall Design:* Traditional STA processes involve particle picking, 3D classification, and refinement steps to solve protein structures. GRIP-Tomo 2.0 aims to offer an alternate approach for identifying proteins in cryo-ET data in a single pass by introducing "cross-domain" classification that incorporates known protein structures from the protein database as a source domain to reduce the need of annotation in a target domain of experimental data. We built on the graph representation techniques established in GRIP-Tomo 1.0 that leverages topological invariant features from evolutionarily conserved protein structures embedded in density maps without realistic noises for protein identification. Here, we further enhanced this program, GRIP-Tomo 2.0 by making it topologically aware of protein structures in realistic noisy backgrounds in a single pass. GRIP-Tomo 2.0 integrates dual-source (i.e., synthetic, experimental) ground truth data generation, topological representation, and interpretable machine learning. The pipeline is designed to extract biologically meaningful patterns from sub-tomograms and leverage them for accurately distinguishing protein densities from realistic imaging noise (**Figure 1, left panel).**

*Ground truth data generation*: GRIP-Tomo 2.0 begins with the ground truth generation of two complementary datasets: synthetic (mock) sub-tomograms simulated from known PDB atomic structures of apoferritin, beta-galactosidase, and aldolase (**Table S1**) under tunable imaging parameters, and experimental sub-tomograms extracted from real cryo-ET data of the mixture from the same three proteins in the mock dataset (see Text S1 and **Figure S1**).

Together, the dual source framework with correlated datasets enables us to synthesize realistic mock sub-tomograms with tunable imaging parameters while anchoring our evaluation in real experimental data for cross-domain classification. The mock dataset provides an essential and tunable source for model training, while the experimental set allows us to evaluate the capacity and biological relevance of learned representations. The ability to calibrate mock volumes to match real tomograms lays a strong foundation for the hybrid validation strategy employed throughout GRIP-Tomo 2.0. **(Figure 1, middle panel).**

*Topological representation of protein structures and interpretable machine learning*: Following data generation, sub-tomograms are converted into graph-based representations that capture evolutionarily conserved topological features of protein structures. This transformation allows GRIP-Tomo 2.0 to encode structural information in a format amenable to scalable computation and statistical learning. The graph construction and feature extraction steps are implemented in a high-throughput, high-performance-computing (HPC)-compatible pipeline capable of analyzing thousands of sub-tomograms in parallel. Finally, GRIP-Tomo 2.0 employs explainable machine learning to classify protein structures and uncover which topological features are most predictive. This integration of interpretable modeling provides fast, and mechanistic insights into what distinguishes protein densities from background noise or from one another in tomograms (**Figure 1, right panel)**.

Workflow validation through particle-picking and benchmarking: To explore GRIP-Tomo 2.0's integration into traditional cryo-ET workflows, we assessed its performance in object detection (i.e., particle-picking) tasks and evaluated RELION[31] classification as a comparison. For particle picking, we used an unseen region of the experimental protein mixture tomogram which had crowded proteins. A sliding-window approach to crop out 64 overlapping subvolumes from that tomogram region and manually labeled them as 'protein' or 'background noise'. GRIP-Tomo 2.0 achieved ~76% binary classification accuracy when tested on these subvolumes. For more details see **Text S2** and **Figure S2**. Additionally, RELION was tested using comparable particle picks and showed limitations when working with small particle pools, struggling to separate the distinct particle classes **(Text S3)**. These findings suggest that GRIP-Tomo 2.0 is versatile and could integrate into broader cryo-ET workflows by performing tasks like particle picking, potentially offering advantages over conventional tools in challenging scenarios, such as those involving low particle counts.

## 2. Mock sub-tomograms generated by tunable pipeline are well-matched to experimental sub-tomograms

*Mock sub-tomogram*: The mock data pipeline (**Figure 2A** and **S1A**) begins with atomic structures from the Protein Data Bank (PDB) from **Table S1** that are input into cisTEM [32] to simulate 2D tilt series projections. The 2D tilt-images are then preprocessed by densmatch [33], Topaz 2D [34] denoising, Contrast Transfer Function (CTF) correction and finally reconstructed into 3D sub-tomograms which go through a low-pass filter and density inversion. A detailed description of the entire pipeline is in **Methods 2.1**. The simulation allows precise control over critical imaging parameters, including the synthesized electron dosage per tilt and synthesized sample thickness, which were the two most impactful factors on the output. The terms 'dosage' and 'thickness' and the correspondent symbols $D$ and $z$ will be used to describe the electron dosage per tilt and sample thickness in the following texts to avoid redundant descriptions. By adjusting $D$ and $z$, we generated mock sub-tomograms that span a wide signal-to-noise ratio (SNR) spectrum—ranging from idealized conditions (**Figure 2C**, first row in golden box) to those mimicking real experimental limitations (**Figure 2C**, grey box). Notably, this mock data generation pipeline is exhibited on HPC platforms and encoded with parallel computing, allowing large-scale simulations across various combinations of imaging parameters.

*Experimental sub-tomogram*: The experimental data pipeline (**Figure 2B** and **S1B)** starts from the tilt series images of a protein mixture solution comprised of the three proteins in **Table S1** collected by the Thermo-Fischer Krios Microscope, as described in **Methods 2.2**, followed by processing such as Contrast Transfer Function (CTF) correction, Topaz 2D denoising and the following 3D reconstruction into a tomogram. We then manually picked the center coordinates of particles and extracted the sub-tomograms based on the coordinates. The experimental data pipeline serves to generate ground truth data from the real world, mainly used as reference and testing set in later steps. Like the mock data, the experimental data generation pipeline also undergoes the deep learning-based Topaz denoising [34] on the 2D tilt series images.

*Imaging parameter sweep*: We first used the same $D$ and $z$ (2.9 e$^-$/Å², 250 Å) as the experimental imaging conditions to simulate the mock sub-tomograms. However, the output sub-tomogram (**Figure 2C**, first row in golden box) retained higher contrast and structural clarity, indicating that the noise and distortion levels were shifted versus the real experimental imaging conditions (**Figure 2C**, grey box). We calibrate the imaging parameters that best reproduce the experimental sub-tomograms (**Figure S3**) by adjusting the simulated $D$ and $z$. We found that mock sub-tomograms simulated at $D$ = 0.3 e$^-$/Å² and $z$ = 500 Å were well-matched to the experimental data in real space by comparing the similarity scores between the two graph representations of the images using several network theory order parameters in GRIP-Tomo 1.0[17]. This calibrated simulation condition was selected for downstream training of the machine learning models, ensuring representational alignment between synthetic and experimental domains for cross-domain classification.

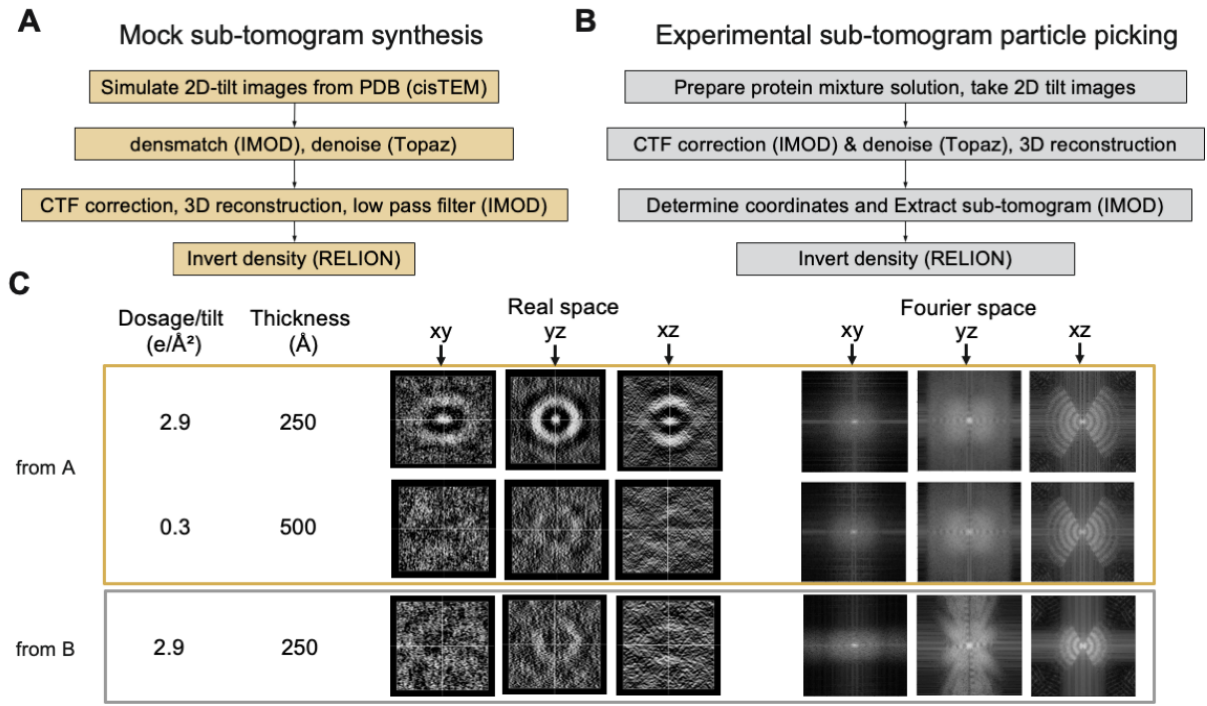

**Figure 2. Synthetic mock and correlative experimental sub-tomograms.**
**(A)** The mock data synthesis pipeline begins with cisTEM simulations to generate tilt-series data from PDB structures from Table S1, followed by densmatch, denoising, contrast transfer function (CTF) correction, and 3D reconstruction. **(B)** The correlative experimental sub-tomogram pipeline begins with sample preparation of protein mixture solution from Table S1, followed by the preprocessing of 2D tilt images, 3D reconstruction of tomogram and particle picking from the tomogram. **(C)** Examples of mock and correlative experimental sub-tomograms. The synthesized mock sub-tomograms exhibit tunable quality by adjusting imaging parameters, such as dosage per tilt (*D*) and sample thickness (*z*), in golden box. The correlative experimental sub-tomogram is shown in grey box with comparable *D* and *z*. Please see the Method section for details.

3. **GRIP-Tomo 2.0 extracts geometry-preserving graph representations from 3D sub-tomograms using scalable HPC workflows**

   *Preprocessing and clustering of density volumes*: The output of ground truth data generation from the last section are sub-tomograms as 3D density volumes stored in MRC format, a common format for storing image and volume data in fields of cryo-electron microscopy and tomography. Each input MRC volume is a 219 × 219 × 219 voxel cube, resulting in approximately 10 million voxels per sub-tomogram. The unit length of a voxel cube is approximately one angstrom for this study. Maintaining sufficient throughput and resolution at this voxel count is essential for preserving meaningful structural information but also poses significant computational challenges for downstream processing at scale. To address this, we developed a streamlined pipeline that reduces the data dimensionality while preserving key topological and structural signals. The input sub-tomogram with $N_{vox}$ ~ 10 million is standardized (**Figure 3A**), thresholded (**Figure 3B–C**), density-aware coarsened (**Figure 3D**) and HDBSCAN[35] clustered (**Figure 3E**) to get a final number of voxels in the largest cluster. For apoferritin, we find that the largest cluster has a size of $N_C$ = 42,029 voxels — closely matching its known number of atoms of approximately 39,000 — and visually recapitulates its spherical geometry even in the presence of tomographic artifacts like the missing wedge effect (**Figure 3E**). This validates that the voxel-to-cluster transformation retains biologically relevant shape information and atom count.

*Graph representation and persistent feature calculation*: From these clusters, we construct undirected graphs from the remaining voxels based on proximity, applying a range of spatial cutoff distances $d'_{cut}$ to capture both local and global connectivity (**Figure 3F**). The increase of $d'_{cut}$ leads to an increased number of edges while the number of nodes is conserved (**Figure 3F**). Graphs generated at shorter $d'_{cut}$ reveal local coordination, while those at longer $d'_{cut}$ capture broader organizational features such as domains or global shape. The resulting graphs preserve the connectivity and the voids of a hierarchical macromolecular structure, including spatial organization affected by missing wedge anisotropy, confirming their geometric fidelity. To quantify these topologies, we extract a set of graph features[36] across scales, such as maximum eigenvector centrality, degree assortativity, and number of communities (**Figure 3G** and **Methods Section 3.1**) into a vectorized "fingerprint". These graph descriptors encode key structural properties and enable machine learning models to distinguish between molecular identities or structural states based on interpretable topological signatures.

*Parallelized and scaled pipeline on High Performance Computing (HPC) platforms*: To further enhance the throughput and scalability of GRIP-Tomo 2.0 to compute graphical features, we deployed a full pipeline on the NERSC Perlmutter supercomputer using Parsl[37] for workflow parallelization (**Figure 3H**). This deployment enabled concurrent processing of 4,096 feature calculations across 1,024 compute nodes with 4 workers per node, achieving an estimated 90-fold speedup compared to serial execution. As a benchmark, the full pipeline—including voxel normalization, density thresholding, HDBSCAN clustering, graph construction, and feature extraction—generated 45,612 graph features from mock datasets in just over two days, consuming approximately 4,323 node hours. This efficient distributed framework allows GRIP-Tomo 2.0 to perform parameter sweeps and process over hundreds of subtomogram sets in matter of days, instead of months.

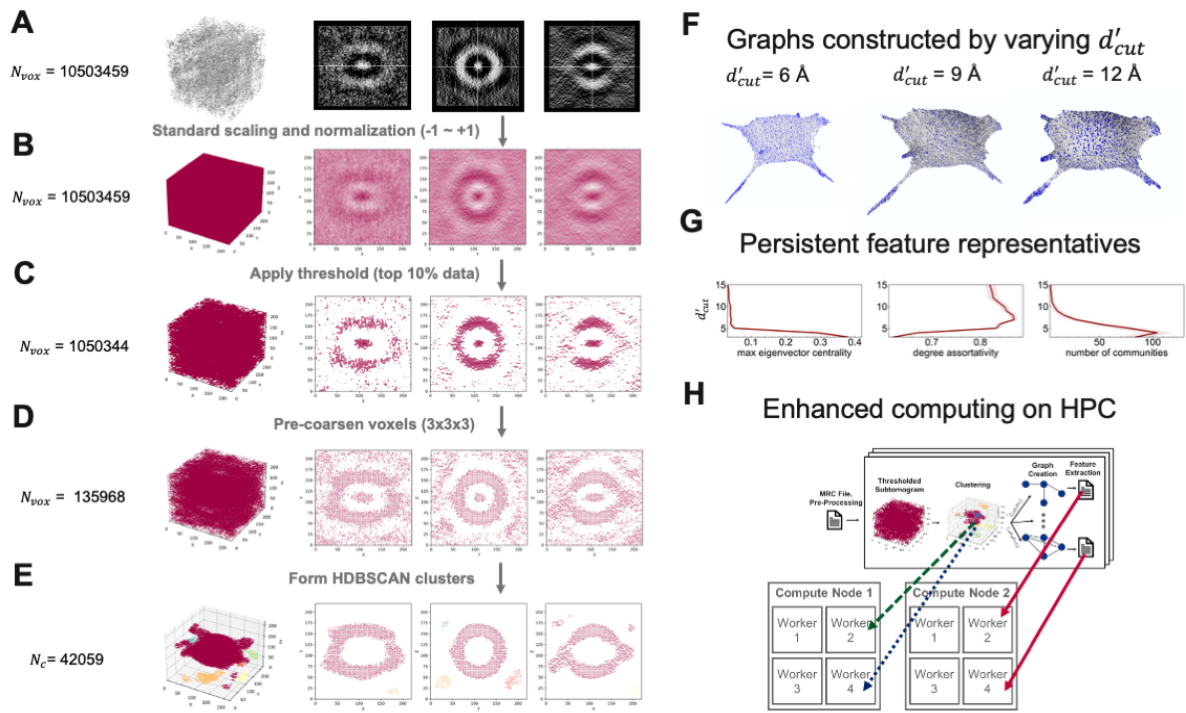

**Figure 3. Extract persistent topological features from the graph representations transformed from a subtomogram using high performance computing.** The GRIP-Tomo 2.0 pipeline transforms high-dimensional sub-tomograms into geometry-preserving graph features for topologically aware protein identification. $N_{vox}$ stands for the number of voxels preserved in each step. We started with $N_{vox}$ =219x219x219. We ended with $N_C$ that stands for the final clustered number of centroids staged for graph constructions. The full HPC pipeline includes the following: **(A)** Standardize raw density maps within the value -1 and 1. **(B–C)** Emphasize structural signals by retaining a certain portion of the highest-density voxels. **(D)** Coarsen the voxels to reduce the node count while maintaining spatial integrity. **(E)** Identify stable point clusters into centroids with HDBSCAN. **(F)** Construct multiple graphs from clustered centroids with varying cutoff distances $d'_{cut}$. **(G)** Extract persistent topological features to capture organizational hierarchy and spatial layout. **(H)** Accelerate calculations with high performance parallel computing.

4. **Topological fingerprints expose imaging condition–dependent separability and reveal classification boundaries of GRIP-Tomo 2.0**

*Diagnostic use of topological fingerprints to guide imaging parameter selection in mock datasets*:

To evaluate how imaging conditions affect GRIP-Tomo 2.0's ability to extract discriminative structural features, we systematically simulated mock sub-tomograms across a matrix of *D* and *z*. From each dataset, we extracted the combined graph-based fingerprints—a multiscale trajectory of 11 topological features across 13 graph cutoff values $d'_{cut}$, —for four categories: horse-l-apoferritin, beta-galactosidase, aldolase, and noise. These fingerprints, visualized in **Figure 4A**. Importantly, some imaging conditions clearly produce fingerprints that are highly separable across all four categories, making them favorable for downstream classification. A particularly illustrative example is the condition with $D = 2.0$ e$^-$/Å² and $z = 400$ Å, where the fingerprint trajectories are visibly distinct even to the human eye. Under this condition, each category exhibits a unique feature profile across $d'_{cut}$, suggesting that the graph topology of these simulated volumes preserves sufficient biological and geometrical signal to support accurate classification—not only between proteins and noise but also among protein classes.

In contrast, other conditions result in overlapping or ambiguous fingerprints. For instance, under $D = 0.3$ e$^-$/Å² and $z = 500$ Å—the condition later selected for simulation-to-experiment domain adapation—the feature trajectories for apoferritin, beta-galactosidase, and aldolase show considerable overlap. While this condition was chosen for its close match to experimental fingerprints, its intrinsic limitations, namely the poor class separability among proteins, make it difficult to distinguish structurally similar macromolecules. As a result, while GRIP-Tomo 2.0 performs well in binary classification (e.g., protein vs. noise in **Figure 5D**) under this condition, it struggles to resolve finer structural differences between proteins like apoferritin and beta-galactosidase (**SI Figure S4**). This observation helps explain the drop in protein-class classification performance seen in the next section: even though the synthetic data were calibrated to match experimental imaging conditions, the condition itself does not support strong topological contrast between proteins. If experimental data were acquired under a more favorable condition—such as higher dosage and lower thickness, yielding fingerprints like those at $D = 2.0$ e$^-$/Å² and $z = 400$ Å—then finer distinctions among proteins would likely be more achievable. This highlights an important dual role of GRIP-Tomo 2.0 fingerprints: not only do they serve as model input, but they also act as diagnostic visual tools that can help researchers pre-screen synthetic imaging conditions and anticipate the success or limitations of downstream classification. In this way, GRIP-Tomo 2.0 offers a novel lens for experiment design and simulation tuning, providing actionable guidance on whether a given imaging regime is sufficient to separate relevant biological classes.

*The classification boundary on mock data revealed by varied sample thickness:*

To further quantify how sample thickness ($z$) affects classification ability, **Figures 4B** and **4C** show F1-scores for GRIP-Tomo 2.0 models trained and tested on mock data generated at fixed $D$ = 0.3 and 2.9 e$^-$/Å² with varying $z$. In general, classification F1-score decreases monotonically with thickness $z$, reflecting increased scattering and loss of structural signal. However, the decline is not uniform across classes, noise-only samples attain higher F1-score than other categories across thicknesses, possibly because of their sparse, irregular graphs differ consistently from protein-like motifs. For $D$ = 0.3 e$^-$/Å² which is the lowest dosage per tilt that is tested, the multi-class classification only has high F1-score when $z$ = 250 Å, indicating most mock samples fall into the correct predicted category. For $D$ = 2.9 e$^-$/Å² which is the highest dosage per tilt that is tested though, both thickness $z$ = 250 and 500 Å yielded multi-class classification with high F1-score. For thickness $z \geq 500$ Å when $D$ = 0.3 e$^-$/Å² and thickness $z \geq 750$ Å when $D$ = 2.9 e$^-$/Å², GRIP-Tomo 2.0's multi-class classification capability drops significantly compared to lower thickness case, highlighting a natural classification boundary that is pushed to higher thickness by higher dosage per tilt D for GRIP-Tomo 2.0.

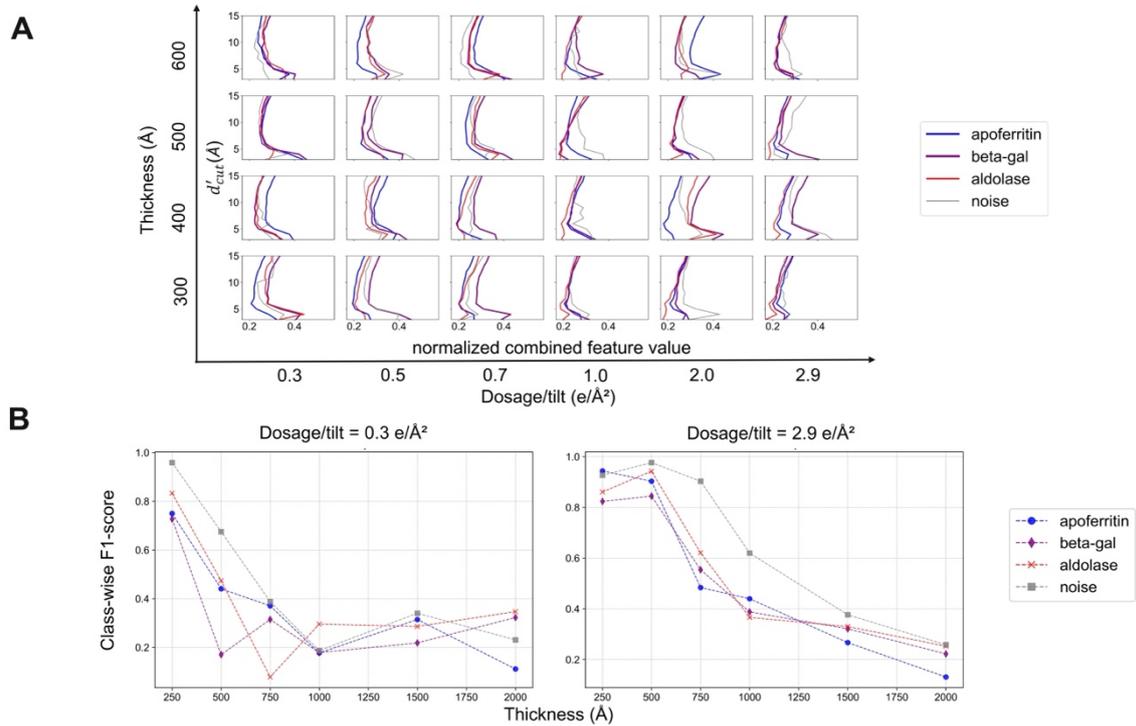

**Figure 4. GRIP-Tomo 2.0 fingerprints identify favorable imaging conditions and reveal classification boundaries across sample thicknesses. (A)** GRIP-Tomo 2.0 feature fingerprints derived from mock sub-tomograms simulated across a grid of electron dosage per tilt $D$ (x-axis, 0.3 to 2.9 e−/Å²) and sample thickness $z$ (y-axis, 300 to 600 Å). Each subplot shows the normalized 13-point feature vector across different $d'_{cut}$, combining 11 topological features described in Methods 3.1.6, and color-coded by structural category: apoferritin (blue), beta-galactosidase (purple), aldolase (red), and noise (gray). **(B)** Class-wise F1-scores varied by thickness $z$ at fixed $D = 0.3$ e−/Å². **(C)** Class-wise F1-scores varied by thickness $z$ at fixed $D = 2.9$ e−/Å².

5. **Calibration of imaging conditions reduces the simulation-to-real shift and enables interpretable topologically aware cross-domain classification for protein density from background noise**

*Calibration of mock sub-tomograms to experimental sub-tomograms*: To bridge the shift between synthetic and experimental sub-tomograms, we investigated whether tuning mock imaging parameters could align topological fingerprints and improve classification performance. Initial mock sub-tomograms were synthesized with imaging parameters in cisTEM ($D = 2.9$ e−/Å², $z = 250$ Å) matching estimated experimental settings. However, when visualizing their topological fingerprints (**Figure S5A**), we observed consistent divergence from experimental fingerprints (**Figure 5A**) across several key features, including average clustering, eigenvector centrality, and degree assortativity shown in the figure. These discrepancies reflect a simulation-to-real shift: although imaging parameters appear nominally matched, the effective signal structures differ. To reduce this domain shift, we performed a calibration step, systematically varying mock imaging parameters (see **Figure 4**, **Figure S3** and **Figure S4**) and identifying the setting—$D = 0.3$ e$^-$/Å$^2$, $z = 500$ Å—that minimized the fingerprint discrepancy. The topological fingerprints of the experimental sub-tomograms and the composite similarity scores [38] between the mock and sub-tomograms are shown in **Figure S4** to support the calibration, along with visual confirmation of the real space images. Using this calibrated condition, we regenerated mock sub-tomograms for three categories (apoferritin, beta-galactosidase, and noise-only), each with 60 rotated instances, yielding a balanced training set of 180 samples. The experimental test set comprised 180 sub-tomograms from the experimental correlated dataset (**Methods 2.2**).

*Improved cross-domain classification after calibration*: We then extracted graph-based persistent features as topological fingerprints from both the training and testing datasets using the GRIP-Tomo 2.0 pipeline and trained a Random Forest classifier. Comparing classification results before (**Figure S5B**) and after (**Figure 5C**) calibration highlights a clear performance gain: uncalibrated training yields poor generalization to experimental data (especially for protein class) with an accuracy of 0.39, whereas calibration recovers strong classification capability, achieving 97% recall for proteins and an overall accuracy of 0.81 (**Figure S5B** and **5C**). The definition of accuracy is provided in **Method 4.2**.

*Interpretability of persistent features and relation to biological scales*: To interpret this performance, we examined the feature importances from the trained classifier (**Figure 5D**). Interestingly, the top-ranked features cluster into three scale bands:

1. Local-range (cutoff 3 Å): capturing local density and node connectivity, potentially related to tightly packed atomic centers or helix cores.

2. Intermediate-range (cutoff 6–8 Å): these cutoffs consistently appeared in earlier GRIP-Tomo1.0 work and may correspond to typical intra-domain distances, such as those spanning α-helices, loops, or β-sheet spacing.

3. Global-range (cutoff 13–14 Å): features like degree assortativity and max centrality at these scales likely reflect global shape, modularity, or domain interfaces. These may correspond to the diameters of the four-fold symmetry channel in apoferritin and central cavity in beta-galactosidase.

This multi-range distribution suggests that GRIP-Tomo 2.0 captures biologically meaningful evolutionarily conserved motifs and hierarchical structures through its graph-based representation. This persistent topological features support the hypothesis that different cutoffs between nodes on a graph representation of protein structure capture structural hierarchy. For instance, the importance of average clustering at cutoff 6 Å may reflect the compactness of secondary structure motifs, while degree assortativity at cutoff 13 Å may relate to global organization patterns such as ion channel diameter or central symmetric domains. Moreover, the calibration strategy demonstrates a practical path to aligning synthetic and real data—a crucial step for enabling cross-domain classification in tomograms. Together, these results validate both the interpretable nature of GRIP-Tomo 2.0 features and the effectiveness of imaging condition calibration in closing the simulation-to-experiment shift for downstream classification.

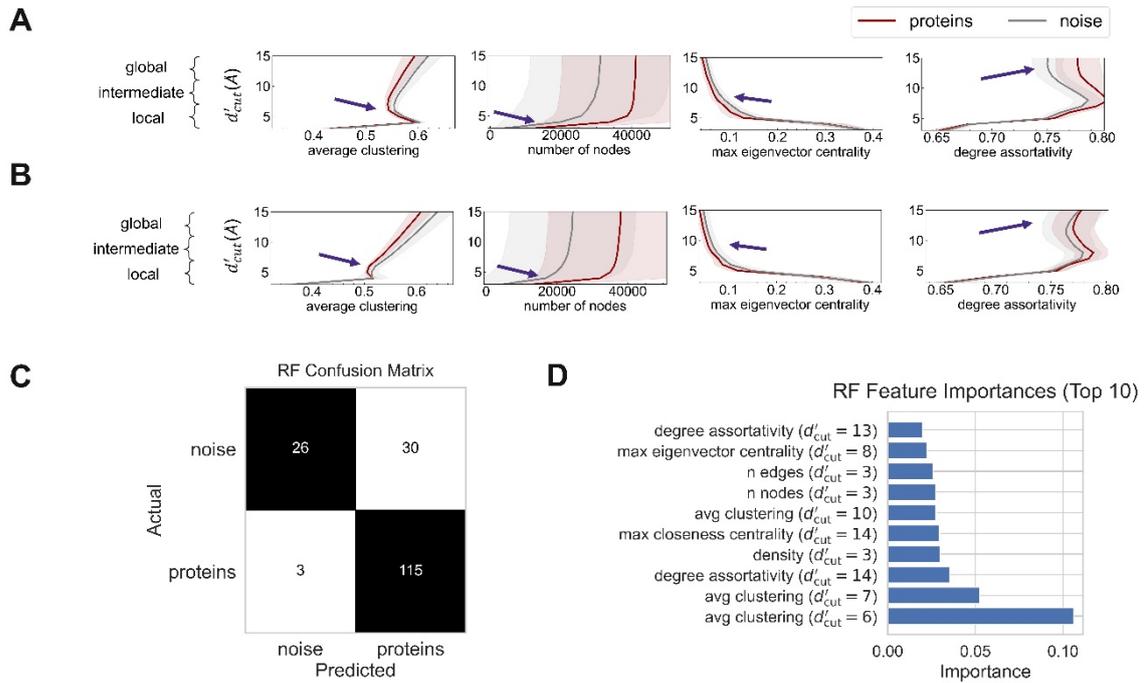

**Figure 5. Cross-domain classification performance.**
**(A)** The source domain: Persistent topological fingerprints from mock sub-tomograms synthesized under calibrated imaging conditions ($D = 0.3$ e$^-$/Å², $z = 500$ Å) as a source domain, showing the graph features that vary by cutoffs $d'_{cut}$: average clustering, number of nodes, max eigenvector centrality, and degree assortativity. Gray and red lines represent noise and protein categories, respectively. **(B)** The target domain: Persistent topological fingerprints of correlated experimental sub-tomograms as a target domain. **(C)** Topologically aware cross-domain classification for proteins: Confusion matrix of cross-domain classification with a model that learned from (A) and tested on (B). **(D)** Top 10 feature importances from the explainable cross-domain classification in (C).

# Discussion

## The evolutionary conserved protein structures provide fingerprints to enable topologically aware identification for protein densities in noisy tomograms

Traditional density-based approaches for cryo-electron tomography (cryo-ET) classification rely heavily on voxel intensities or template correlations, making them vulnerable to variations in signal-to-noise ratio, missing-wedge artifacts, and imaging parameters. In contrast, GRIP-Tomo 2.0 introduces a topology-driven alternative that leverages structural connectivity, holes, and voids [18] between secondary structures or domains, rather than raw density as the primary descriptor of molecular identity. By encoding macromolecular architecture as graphs that preserve the fundamental relational geometry of a protein, the framework captures evolutionarily conserved topological signatures—features that are not strongly dependent on the exact details of atomistic structures. The qualitative features of spatial structures remain invariant under rotation, translation, and even local deformation due to missing wedge. This conceptual shift transforms protein identification from an intensity-matching problem into one grounded in the intrinsic organization of biological form as demonstrated in GRIP-Tomo 1.0[17], enabling recognition that transcends imaging artifacts or noise as shown in the current work.

Cryo-ET enables visualization of macromolecules within their native environments, yet its inherently low signal-to-noise ratio continues to impede reliable classification. GRIP-Tomo 2.0 addresses this limitation by extracting persistent topological patterns that remain stable across replicates and imaging conditions. Representing three-dimensional structures as graphs of connectivity allows the model to capture the hierarchical representations and the voids or cavities that endure even when voxel intensities are degraded [39] in complex multidimensional datasets. These graph-based fingerprints emphasize the shape, or the geometric backbone of macromolecular organization that persists across different scales to distinguish robust signal from noise, enabling effective identification of volumes containing protein signatures without segmentation or template matching. In doing so, GRIP-Tomo 2.0 provides a topologically informed and biologically interpretable representation space, bridging structural conservation with robust machine-learning-based identification of protein density from noise according to the nuances in protein topology.

**Graph representation of proteins enables single-pass, cross-domain protein classification from noise with efficiency and interpretability**

GRIP-Tomo 2.0 advances protein identification from background noise beyond voxel-intensity learning by enabling single-pass, cross-domain classification from synthetic to experimental cryo-ET data. The framework trains exclusively on mock sub-tomograms generated from atomic models with realistic imaging conditions [32], each representing a unique structural topology derived from a single known conformation. Despite this restricted structural diversity, the model learns topological fingerprints that remain stable under varying signal-to-noise ratios and missing-wedge effects. When applied directly to experimental data without retraining, these fingerprints retain discriminative power, demonstrating genuine domain transferability between synthetic and real tomograms [22, 40]. By emphasizing the structural connectivity and the voids between hierarchical structures of macromolecules rather than their density amplitudes, GRIP-Tomo 2.0 alleviates the dependence on large, labeled datasets—an enduring limitation in cryo-ET recent techniques [8, 41-46] —and alternatively provides a structurally grounded route to generalizable learning across data modalities, unlocking the full potential to discover the density of proteins, large or small, from noise over the entire tomogram in a single pass for science discovery.

Beyond accurate classification, GRIP-Tomo 2.0 offers an interpretable bridge between data quality and biological insight through topological feature visualization. By systematically varying imaging parameters, the model reveals how persistent features deform across the synthetic-to-experimental continuum, pinpointing optimal signal to noise ratio (SNR) and thickness conditions that balance realism with discriminability. Remarkably, maximal alignment between mock and experimental fingerprints occurs not under the highest image quality, but at intermediate electron dosage (D ≈ 0.3 e$^-$ Å$^{-2}$) and moderate thickness (z ≈ 500 Å)—conditions that best reproduce experimental variability. These results highlight that realism and task alignment, rather than maximal contrast, are essential for cross-domain generalization. Furthermore, to validate the approach, we implemented a sliding-window particle-picking test on unseen regions of the experimental tomogram. Using 64 sub-volumes cropped from a 5813 Å$^3$ region, GRIP-Tomo 2.0 achieved a binary accuracy of ~76% in distinguishing proteins from background noise, validating that the workflow can perform single-pass detection and classification within complex tomograms. This experiment underscores the scalability of topological fingerprints for both identification and localization tasks in realistic cryo-ET environments.

**Topological fingerprints enable imaging-aware synthesis design, cross-domain calibration and classification**

A defining strength of GRIP-Tomo 2.0 lies in the dual role of its topological fingerprints: they function not only as model inputs for protein classification from noise but also as diagnostic visual tools that guide experimental and synthetic imaging design. By mapping the separability of topological features across simulated dosage–thickness matrices, researchers can visually identify regions where structural information is preserved despite noise, thereby pre-screening synthetic conditions before costly data generation. This dual capability establishes GRIP-Tomo 2.0 as both a predictive and analytical framework, enabling imaging-aware synthesis design and calibration across the simulation-to-experiment domains [47-49].

The interpretability of GRIP-Tomo 2.0 arises from its graph-theoretic foundation, in which persistent features such as clustering coefficient, assortativity, and eigenvector centrality serve as biologically meaningful descriptors that capture hierarchical molecular structures in different scales. These features not only drive classification accuracy but also correlate with evolutionarily conserved structural motifs, including α-helices, β-sheets, symmetric domains, and the holes or cavities in a molecular complex. The resulting graph fingerprints form a compact and low-dimensional representation of macromolecular architecture, consistent with recent findings that cryo-EM imaging data occupies an intrinsically low-dimensional manifold [50]. Feature-importance analyses of trained models consistently highlight these interpretable descriptors, revealing how cross-domain classification between synthetic and experimental data is achieved. Moreover, the capacity to visualize feature separability across imaging parameters provides researchers with an intuitive means to benchmark data quality—an advantage absent in most black-box deep learning approaches—thus promoting transparent and reproducible cryo-ET workflows.

Another critical advantage of GRIP-Tomo 2.0 is its scalability in synthetic data generation and integration. The pipeline can automatically produce large, annotated subtomogram datasets that are both topologically aware and experimentally calibrated. In contrast to other approaches [51, 52] which typically depend on voxel-based or segmentation-driven representations, GRIP-Tomo 2.0 focuses on single particle subtomogram data with their associated topologically aware and interpretable features, leveraging the knowledge of structural fingerprints in graph representations that are inherently interpretable and transferable. This enables downstream integration with AI models requiring extensive, structured inputs, such as graph embeddings or transformers [46, 53, 54], particularly for challenging tasks involving small proteins or rare conformational states. By reducing dependence on manual labeling and emphasizing structural interpretability, GRIP-Tomo 2.0 functions as both a classification engine and a data-generation platform for the broader cryo-ET community. Nonetheless, feature extraction remains computationally demanding even on HPC systems, underscoring the need for GPU-accelerated implementations to support large-scale adoption in future development of GRIP-Tomo.

**Limitations of GRIP-Tomo 2.0 and future steps**

Despite these strengths, several limitations remain. First, the mock data generation requires manual tuning to match experimental conditions. Future work will automate this process using a data-driven approach. Second, cross-domain classification struggles to distinguish between proteins (e.g., apoferritin vs. beta-galactosidase, see **Figure S6**), likely due to fingerprint overlap under current imaging conditions; improved separation may be possible with higher dosage and thinner samples. Third, while we explored the effect of sample thickness in detail, future work should systematically vary dosage and build a complete condition-performance matrix. Lastly, our method has so far been validated only on a single experimental dataset. Demonstrating its generalizability across other experimental conditions (such as the well-known crowding conditions that impede cryo-ET particle picking[46]) will be an important next step, for example, to explore integrating our approach by leveraging the structure prediction driven by artificial intelligence (AI) combined with subtomogram analysis in understanding macromolecular assemblies [13] as part of a visual proteomics workflow [55].

**Conclusion**

To conclude, GRIP-Tomo 2.0 expands the computational toolset for cryo-electron tomography by providing a novel solution for particle identification workflows. Building upon our previous framework, this enhanced platform integrates tunable mock data simulation incorporating realistic noise, and a novel interpretable feature extraction pipeline with high-performance computing acceleration. The key innovation lies in extracting biologically relevant molecular "fingerprints" that enable effective cross-domain learning from simulated to experimental data while requiring minimal training datasets.

# Materials and Methods

1. **Graph based analysis of density volumes**

In our previous work, we developed GRaph Identification of Proteins in Tomograms (GRIP-Tomo) to identify proteins in pristine synthetic volume densities [17] in one pass, illustrated in Figure 6 of [17]. Briefly, in this method, an input sub-volume is normalized and thresholded. The remaining high-density voxels above the threshold are clustered using DBSCAN [56]. The cluster centroids as assigned as nodes, and edges are added between the nodes if their Euclidean distance is below a cutoff value, $d_{cut}$ and $d'_{cut}$ for PDB structure and density volumes. From this graph we calculate a vector of 12 topological graph features: number of nodes, number of edges, number of communities, density, average clustering of each node, degree assortativity, diameter, average paths length, clique number, max betweenness centrality, max closeness centrality, max eigenvector centrality. The average relative similarity can be computed from two graph feature vectors. We used a single short cutoff of 8 or 9 angstroms to distinguish single-domain and multiple-domain proteins.

In GRIP-Tomo 2.0, we build on our previous work, creating a mathematical graph representation $G = (V, E)$ using the HDBSCAN algorithm [35, 57] instead of DBSCAN. The set of nodes V correspond to the cluster centroids from HDBSCAN, and the set of edges E accounts for the relations between nodes. An edge exists between any two nodes if they are within a cutoff distance $d'_{cut}$. The distance between two nodes is the minimum Euclidean distance. Thus, the mathematical graph representation is defined as an adjacency matrix, where $d'_{cut}$ is an important hyperparameter regulating the connectivity of the graph.

## 2. Ground Truth Data Generation

### 2.1 Synthesis of single-particle mock sub-tomograms from known protein structures

For the synthesis of a mock sub-tomogram with a protein and artifacts (**Figure 2A** and **S1A**), we used models available at the Protein Data Bank (PDB) [58] as input structures to cisTEM [32] to generate a simulated series of 2D images with defined tilt angle, noise ratio and experimental artifacts. Specifically, we used models of horse spleen light chain apoferritin (PDB ID: 2W0O), $\beta$-galactosidase (PDB ID: 6DRV), and aldolase (PDB ID:8EW2) **(SI Table S1)**. We simulated combinations of artifacts by varying the electron dosage per tilt frame ($D$) and sample thickness ($z$). All 2D tilt series were simulated from -60 to 60 degrees with 3-degree increments.

To ensure interoperability, the simulated 2D tilt images were density matched to an experimental reference dataset (see Methods 2.2) using IMOD [33] before applying Topaz 2D denoising [34] using the affine model with a patch size of 64. The denoised tilt series was merged together using EMAN2 [59] and then contract transfer function (CTF) corrected, reconstructed, and low pass filtered using IMOD. The mock sub-tomograms density was inverted using RELION 4 [31] to harmonize the simulated and experimental datasets.

We also synthesized noise-only sub-tomograms lacking protein density as a control data set. These were simulated using the same process as those described above, except that the starting protein structure was replaced by a single carbon atom at the center of the space.

### 2.2 Collecting single-particle experimental sub-tomograms

To validate our approach, we extracted sub-tomograms from an experimental tomogram (**Figure 2B** and **S1B**), collected by the Thermo-Fischer Krios G3i microscope at the Enviornmental Molecular Sciences Laboratory (EMSL). The experimental tomogram contains a mixture of three proteins that are structurally discernable in size and shape: equine spleen light chain apoferritin (Sigma Aldrich, #A3641), *E. coli β*-galactosidase (Sigma Aldrich, #G5635) and rabbit aldolase (Sigma Aldrich, #A8811). Apoferritin and *β*-galactosidase have different helical and beta-sheet composition and different oligomeric assembly and symmetry but they have a common total molecular weight ~450-460kDa. Aldolase on the other hand shares the same oligomeric assembly and symmetry as *β*-galactosidase but is roughly 1/3 the total molecular weight at ~160kDa. This combination then allows testing proteins of similar size but different composition and shape, or similar shape and composition. This solution was made of a sparse mixture of each protein at an individual concentration of 1.5mg/mL in a buffer of 25mM Tris, 2mM MgCl2, 50mM NaCl, and 2mM DTT. 3uL of the protein mixture was loaded onto a 1.2/1.3 holey carbon TEM grid, blotted under 90% RH, and vitrified by plunge freezing into liquid ethane. This solution of protein mixtures has an estimated location-dependent thickness of 200~300 Å, as the ice layer formed a meniscus shape in the holes that was thinnest in the center and thickest near the edge. We used the average value of 250 Å as the thickness of this experimental tomogram when performing the synthetic data calibration.

Screening and tilt series collection was performed with serialEM [60] in a dose symmetric strategy with images every 3 degrees to +/- 60 degrees, yielding total 41 tilt images each with a dose of 2.9 e/Å$^2$. The total cumulative electron dose for the tilt series was 120e/Å$^2$ at 42,000 times magnification for a pixel size of 1.1 Å/pixel and a 20eV energy filter slit. Raw frames were motion corrected with Motioncor2 [61], denoised with Topaz using the affine model with a patch size of 1024 [34], and the tilt series were aligned and reconstructed in IMOD using weighted back projection. CTF fitting and correction was performed with IMOD's CTF plotter prior to reconstruction. Tomograms were visualized with the 3dmod module in IMOD for particle picking, and the subvolumes were extracted using the 'trimvol' command in IMOD [33].

From the reconstructed tomogram, we manually picked 180 sub-volumes split across each of the three categories: apoferritin, beta-galactosidase and noise-only. Due to its small size, aldolase was not picked as ambiguous. Sub-tomograms centered on the picked particles were extracted using IMOD [33] and inverted with RELION 4 [31]. For more detailed information on the mock and experimental data preparation, see **SI Text S1**.

## 3. HPC enhanced workflow of extracting topological features from mock and experimental density volumes

### 3.1 Density volumes to features

To extract interpretable graph-based features from tomographic data, we developed a high-throughput workflow that transforms 3D density volumes (in MRC format) into topological descriptors. This pipeline was optimized for HPC deployment using Parsl [37] and executed on ALCC Perlmutter.

Standard scaling and normalization: We applied per-volume standard scaling to ensure consistency in density distribution across MRC files. Using scikit-learn's StandardScaler [62], voxel intensities were rescaled via z-score normalization: $z=(x-\mu)/\sigma$, where $\mu$ and $\sigma$ denote the mean and standard deviation of the MRC volume. This step enabled fair comparison between mock and experimental datasets, which often exhibit different dynamic ranges and baseline densities.

Thresholding: A threshold ratioing step retained the top 10% highest-density voxels in each standardized volume. This top-k percentile filtering caps the number of candidate voxels passed downstream, bounding the computational complexity and memory usage of graph construction. The ratio was chosen to maximize feature extraction completeness while ensuring the pipeline could run within time and memory constraints on HPC systems.

Density coarsening: We introduced a density-aware voxel coarsening strategy with NumPy [63] to reduce spatial and computational load. Voxels were grouped in 3×3×3 windows, and their average position and density were computed to generate a representative centroid. This strategy preserved local density peaks and reduced worst-case complexity up to $1/C_{size}^3$ for cube size $C$, supporting scalability even for volumes with >10 million voxels.

HDBSCAN (Hierarchical Density-Based Spatial Clustering of Applications with Noise) clustering[35]: To isolate protein signals from background noise, the coarsened point cloud was clustered using HDBSCAN. Unlike DBSCAN, which is sensitive to parameter tuning at low thresholds, HDBSCAN identifies stable clusters over a range of densities, enabling robust segmentation. The largest persistent cluster from each volume was selected as the candidate region for downstream graph representation. Clustering parameters were tuned to perform well across synthetic and experimental datasets (e.g., min_samples=512, min_cluster_size=512).

Graph construction: From the retained cluster points, graphs were constructed across multiple spatial resolutions. Each graph was built by assigning edges between node pairs within a cutoff distance $d'_{cut}$. We varied $d'_{cut}$ from 3 Å to 15 Å at a 1-angstrom increment to capture persistent structural features across scales, with larger cutoffs either running out of memory or time. The resulting multi-scale graphs reflect both local density geometry and global topological connectivity.

Persistent graph feature calculation across $\mathbf{d'_{cut}}$: For each graph at every cutoff level, a comprehensive set of 11 graph features was extracted, including degree assortativity, clustering coefficient, eigenvector centrality, closeness centrality, and others. This set of features across every $d'_{cut}$ were aggregated into a single vector representation with a size of 11 graph features x 13 $d'_{cut}$ values = 143 machine learning features per sub-tomogram and saved as comma-separated valued (CSVs) for subsequent machine learning (ML) analysis. To improve scalability, we excluded features with poor runtime efficiency, such as max clique number, which we identified as a key bottleneck during profiling.

By combining these features across biological relevant multiple scales, we created a persistent graph feature vector of the data as a foundation for training explainable machine learning models. These persistent feature vectors are then passed into a Random Forest (RF) [64] model to perform model training and testing, as described in the Methods section 4.

### 3.2 HPC Parallelization Using Parsl

To support scalable feature extraction from thousands of sub-tomograms under various parameter settings, we deployed the pipeline on the NERSC Perlmutter supercomputer using Parsl, a Python-based parallel scripting library[37]. Parsl's dataflow-aware scheduler orchestrates concurrent task execution while preserving data dependencies between stages. Each processing step—standardization, thresholding, clustering, graph construction, and feature extraction—was parallelized across independent sub-tomograms or parameter combinations.

On Perlmutter, we were able to use 1024 nodes to process 4096 sub-tomogram volumes in parallel. Parsl allowed different pipeline stages to run asynchronously across workers, dramatically reducing total wall-clock time from a projected 180 days on a single system to 2 days on Perlmutter – a 90-fold speed up. We collected over 100,000 graph feature sets across the mock and experimental datasets, enabling thorough downstream analysis.

Execution time profiling across 45,000 feature set extraction jobs further guided our optimizations, identifying steps like clique number and graph construction as major bottlenecks for large, dense graphs. These findings motivated improvements such as feature pruning and the use of coarsening to reduce graph size.

## 4. Training, Evaluation and Interpretability of the learning model

### 4.1 Machine learning model

To evaluate the discriminative power of GRIP-Tomo 2.0 graph-based features and assess their biological relevance, we trained and interpreted a supervised Random Forest (RF) classifier [64] using Python's scikit-learn [62] implementation and 1000 estimators. RF models are well suited for classification tasks with low data amounts and provide feature importance values for interpretability.

We constructed multi-class classification including aldolase, apoferritin, beta-galactosidase and noise in **Figure** 4, the Random Forest model was trained and tested on the same mock dataset. For each imaging condition with a certain combination of $D$ and $z$, 200 mock sub-tomograms (50 each of aldolase, apoferritin, beta-galactosidase, and synthetic noise) were simulated following the pipeline in **Figure 2A** and **S1A**. These mock sub-tomograms were then randomly split as 10% for training and 90% for testing.

We also constructed a binary classification task to distinguish between protein-containing and noise-only sub-tomograms in **Figure 5**. The training dataset consisted of 180 mock sub-tomograms (60 each of apoferritin, beta-galactosidase, and synthetic noise), generated under the calibrated imaging condition (electron dosage = 0.3 e⁻/Å², thickness = 500 Å). The test set comprised 180 experimental sub-tomograms annotated as protein or noise via manual inspection.

**4.2 Performance evaluation:**

Model performance of the trained Random Forest model was assessed using standard classification metrics derived from the confusion matrix using the scikit-learn package [62] in Python. In confusion matrix, predictions were categorized as true positives (TP), false positives (FP), false negatives (FN), and true negatives (TN).

Overall **accuracy** was computed as (TP+TN)/(TP+FP+FN+TN).

The **F1-score** was defined as 2×(Precision×Recall)/(Precision+Recall), where Precision=TP/(TP+FP and Recall=TP/(TP+FN).

In addition, **class-wise accuracy** was calculated as $TP_i/(TP_i+FN_i)$, where i is a specific category, providing a per-class evaluation beyond the global accuracy.

**4.3: Interpretability of feature importance in a learnt model**

In GRIP-Tomo 2.0, we calculate the importance of each feature at each $d_{cut}'$ by scikit learn feature importance [62]. The feature importance serves as a metric for measuring the sensitivity of the predicted result to the change of inputs. Thus, this approach provides insights into which features contribute most significantly to the model's prediction.

**Software, data, and tools availability**

The GRIP-Tomo source code is available at https://github.com/EMSL-Computing/grip-tomo. The datasets used during this study will be made publicly available upon journal acceptance. Generative AI tool usage: ADEPT (Agentic Discovery and Exploration Platform for Tools) [65] was used to assist with literature retrieval and review. GitHub Copilot assisted with improving code quality (e.g., unit tests, docstrings, refactoring), and ChatGPT (versions 4o and 5) assisted with copyediting and grammar for readability. All AI-generated outputs were supervised and reviewed by the authors.

# Acknowledgements

This work was supported by the U. S. Department of Energy (DOE), Office of Science, Office of Biological and Environmental Research, under the NW-BRaVE Bio-preparedness project (FWP 81832) and the BSSD Krios Operations project (FWP 74915). A portion of this research was performed on a project award (https://doi.org/10.46936/staf.proj.2023.61054/60012367) from the Environmental Molecular Sciences Laboratory, a DOE Office of Science User Facility sponsored by the Biological and Environmental Research program under Contract No. DE-AC05-76RL01830. Pacific Northwest National Laboratory is a multi-program national laboratory operated by Battelle for the DOE under Contract DE-AC05-76RL0 1830. An award of computer time was provided by the ASCR Leadership Computing Challenge (ALCC) program. This research used resources of the National Energy Research Scientific Computing Center (NERSC), a Department of Energy User Facility using NERSC award ALCC-ERCAP0034213.